\documentclass{easychair}

\usepackage[english]{babel}
\usepackage{amsfonts}
\usepackage{inputenc}
\usepackage[ruled]{algorithm}
\usepackage{xcolor}
\usepackage{graphicx}
\usepackage{algpseudocode}
\usepackage{booktabs,multicol,multirow}
\usepackage{fancyhdr}
\usepackage{epic}
\usepackage{xspace}
\usepackage{amssymb,amsmath,verbatim}
\usepackage{multirow}
\usepackage{multicol}
\usepackage{adjustbox}
\usepackage{threeparttable}
\usepackage{float}

\usepackage{amsthm}
\usepackage{chngcntr}
\usepackage{url}
\usepackage[
n,
operators,
advantage,
sets,
adversary,
landau,
probability,
notions,	
logic,
ff,
mm,
primitives,
events,
complexity,
asymptotics,
keys]{cryptocode}
\usepackage{threeparttable, tablefootnote}

\newtheorem{thm}{Theorem}

\newtheorem{defn}[thm]{Definition}

\def\Adv{\mathsf{Adv}}
\def\ZZ{\mathbb{Z}}
\def\RR{\mathbb{R}}
\def\RR{\mathbb{R}}

\def\u{\bf{u}}
\def\v{\bf{v}}
\def\e{\bf{e}}
\def\b{\bf{b}}
\def\m{\bf{m}}
\def\L{\Lambda}

\def\b{\bf{b}}
\def\s{\bf{s}}
\def\A{\bf{A}}
\def\H{\bf{H}}
\def\R{\bf{R}}
\def\G{\bf{G}}
\def\X{\bf{X}}

\def\B{\bf{B}}

\def\x{\bf{x}}
\def\c{\bf{c}}
\def\IBPRE{\mathsf{IB\hbox{-}uPRE}}
\def\SIBPRE{\mathsf{Selective\hbox{-}IB\hbox{-}uPRE}}
\def\AIBPRE{\mathsf{Adaptive\hbox{-}IB\hbox{-}uPRE}}
\def\INDsCPA{\mathsf{IND\hbox{-}sID\hbox{-}CPA}}
\def\INDCPA{\mathsf{IND\hbox{-}ID\hbox{-}CPA}}
\def\Exp{\mathsf{Exp}}
\def\Pr{\mathsf{Pr}}

\title{Collusion-Resistant Identity-based Proxy Re-Encryption:
Lattice-based Constructions in Standard Model}

\author{
Priyanka Dutta\and Willy Susilo\and Dung Hoang Duong\and Partha Sarathi Roy
}

\institute{
  Institute of Cybersecurity and Cryptology\\
School of Computing and Information Technology\\
University of Wollongong\\
	Northfields Avenue, Wollongong NSW 2522, Australia\\
\email{\{pdutta,wsusilo,hduong,partha\}@uow.edu.au}
 }

\authorrunning{}

\titlerunning{}

\begin{document}

\maketitle

\begin{abstract}
The concept of proxy re-encryption (PRE) dates back to the work of Blaze, Bleumer, and Strauss in 1998. PRE offers delegation of decryption rights, i.e., it securely enables the re-encryption of ciphertexts from one key to another, without relying on trusted parties. PRE allows a semi-trusted third party termed as a ``proxy" to securely divert encrypted files of user A (delegator) to user B (delegatee) without revealing any information about the underlying files to the proxy. To eliminate the necessity of having a costly certificate verification process, Green and Ateniese introduced an identity-based PRE (IB-PRE). The potential applicability of IB-PRE sprung up a long line of intensive research from its first instantiation. Unfortunately, till today, there is no collusion-Resistant unidirectional IB-PRE secure in the standard model, which can withstand quantum attack. In this paper, we present the first concrete constructions of collusion-Resistant unidirectional IB-PRE, for both selective and adaptive identity, which are secure in standard model based on the hardness of learning with error problem. 
\end{abstract}




\section{Introduction}
In a Proxy Re-encryption (PRE) scheme, a proxy is given some information that allows turning a ciphertext encrypted under a given public key into one that is encrypted under a different key. A naive way for a user $A$ to have a proxy implementing such a mechanism is to simply store her private key at the proxy: when a ciphertext arrives for $A$, the proxy decrypts it using the stored secret key and re-encrypts the plaintext using $B$’s public key. The obvious problem with this strategy is that the proxy learns the plaintext and $A$’s secret key. Blaze, Bleumer, and Strauss \cite{BBS98} introduced the concept of PRE to achieve an efficient solution that offers delegation of decryption rights without compromising privacy. PRE involves a semi-trusted third party, called a proxy, to securely divert encrypted files of one user (delegator) to another user (delegatee). The proxy, however, cannot learn the underlying message $m$, and thus both parties' privacy can be maintained. PRE (and its variants) have various applications ranging from encrypted email forwarding \cite{BBS98}, securing distributed file systems \cite{AFGH07}, to digital rights management systems \cite{Smith05}. We notice a real-world file system employing a PRE scheme by Toshiba Corporation \cite{MHS13}. On the other hand, various emerging ideas and techniques have shown connections between re-encryption with other cryptographic primitives, such as program obfuscation \cite{HRV07,CCV12,CCLNX14}, and fully-homomorphic encryption \cite{CLNTV15}. Hence, further studies along this line are both important and interesting for theory and practice.

PRE systems are, mainly, classified as unidirectional and bidirectional based on the direction of delegation. It is worth mentioning that bidirectional construction easily implementable using a unidirectional one. The first unidirectional PRE was proposed by Ateniese et al. in \cite{AFGH07}, where following desired properties of a PRE are listed: 
\begin{itemize}
    \item {\em Non-interactivity}: {\em re-encryption key}, $rk _{A \rightarrow B}$, can be generated by $A$ alone using $B$'s public key; no trusted authority is needed; 
    \item {\em Proxy transparency}: neither the delegator nor the delegatees are aware of the presence of a proxy; 
    \item {\em Key optimality}: the size of $B$'s secret key remains constant, regardless of how many delegations he accepts; 
    \item {\em Collusion resilience}: it is computationally infeasible  for the coalition of the proxy and user $B$ to compute $A$’s secret key; 
    \item {\em Non-transitivity}: it should be hard for the proxy to re-delegate the decryption right, namely to compute $rk _{A \rightarrow C}$ from $rk _{A \rightarrow B}$, $rk _{B \rightarrow C}$. 
\end{itemize}
To achieve the aforementioned properties (partially) with improved security guarantee, there are elegant followup works that can be found in \cite{CH07,HRV07,LV08,CCV12,CCLNX14}. Unfortunately, all the aforementioned constructions are vulnerable to quantum attacks. It is a need of the age to construct a quantum-safe version of useful cryptographic primitives like PRE. Gentry \cite{Gentry09} mentioned the feasibility of unidirectional PRE through a fully homomorphic encryption scheme (FHE). However, FHE costs huge computation. Xagawa proposed the construction of PRE in \cite{Xagawa10}, but the construction lacks concrete security analysis. Further development of lattice-based PRE can be found in \cite{KIRSH14,CCLNX14,NX15,FL19}. In \cite{KIRSH14}, the first non-interactive CCA secure lattice-based PRE proposed. Unfortunately, there is some issue regarding security reduction, which is fixed by Fan et al. \cite{FL19}.

The Certificate management problem is one of the most crucial issues in the PKI based schemes. This crucial issue was addressed by Green et al. \cite{GA07} in the area of PRE. For lattice-based construction, Singh et al. \cite{SRB13} proposed a bidirectional identity-based PRE. However, it is required to use the secret key of both delegator and delegatee to generate a re-encryption key, which lacks one of the fundamental properties of PRE. Further, they proposed unidirectional identity-based PRE \cite{SRB14u,SRAS20}, termed as {\em $\IBPRE$}, secure in the random oracle model. However, the size of the re-encrypted ciphertext blows up than the original encrypted one. Thus, \cite{SRB14u,SRAS20} lack the property called {\em Proxy transparency}. Recently, Dutta et al. \cite{DSDBR20} proposed {\em $\IBPRE$} secure in standard model. Unfortunately, {\em $\IBPRE$} of \cite{DSDBR20} is not collusion resistant. There are some further attempts to construct lattice-based identity-based PRE, which are flawed\footnote{In \cite{HJGS18}, authors claimed to prove IND-ID-CPA, but provide the proof for IND-CPA. In \cite{yin17}, authors assumed a universally known entity ({$\G$} matrix; see section \ref{1.lattices}) as a secret entity.} \cite{HJGS18,yin17}.

\noindent{\bf Our Contributions and Techniques:} It is an interesting open research problem to construct post-quantum secure collusion-Resistant $\IBPRE$ in the standard model. In this paper, we resolve this daunting task by constructing concrete schemes based on the hardness of {\em learning with error} (LWE) problem. We propose, both, selectively-secure $\IBPRE$ and an adaptively-secure $\IBPRE$. Proposed schemes enjoy the property of proxy transparency, i.e., a recipient of ciphertext cannot distinguish whether the ciphertext is the original one or re-encrypted. Furthermore, the proposed constructions have properties like non-interactivity, key optimality, non-transitivity along with other properties follow generically from IB-PRE. We provide a comparison among existing (IB)PRE and proposed schemes in Table \ref{T1}. 

{\renewcommand{\arraystretch}{1.2}
	\setlength{\tabcolsep}{2.5pt}
	\begin{table*}[ht]
		\label{T1}
		\begin{center}
			\caption{Comparison with Lattice-based (IB)PRE}
			\small
			\begin{tabular}{c c c c c c }
				\toprule
				Scheme & Unidirectional & PT & CR & IBE & Std.\\
				\midrule
				\cite{ABW13} & $\times$ &  $\checkmark$ & $\times$ & $\times$ & $\checkmark$\\
				\cite{KIRSH14} & $\checkmark$ &  $\checkmark$ & $\checkmark$ & $\times$ & $\checkmark$\\
				\cite{NX15} & $\checkmark$ &  $\times$ & $\checkmark$ & $\times$ & $\checkmark$\\
				\cite{FL19} & $\checkmark$ & $\times$ & $\checkmark$ & $\times$ & $\checkmark$\\
				\cite{SRB14u,SRAS20} & $\checkmark$ & $\times$ & $\checkmark$ & $\checkmark$ & $\times$\\
				\cite{DSDBR20} & $\checkmark$ &  $\checkmark$ & $\times$ & $\checkmark$ & $\checkmark$\\
				Proposed & $\checkmark$ &  $\checkmark$ & $\checkmark$ & $\checkmark$ & $\checkmark$\\
				Schemes &  &  & & & \\
				\toprule
				
				\noalign{\smallskip}
				
				\multicolumn{6}{l}{${}^{\phantom{**}}$ \parbox[t]{0.55\textwidth}{PT: Proxy Transparency; CR: Collusion-Resistant; IBE: Identity-based Encryption; Std.: security in standard model.}}

			\end{tabular}
			\label{tab1}
		\end{center}
		
	\end{table*}
} 

To construct the $\IBPRE$, we start with the construction of the identity-based encryption scheme by Agrawal et al. \cite{ABB10-EuroCrypt}. From very high level, it may seems that the security reduction works as in \cite{ABB10-EuroCrypt}. But, challenges arise during the simulation of ReKey oracle. We explain the devised techniques to combat such challenges in subsequent paragraphs.

For selectively-secure $\IBPRE$ scheme, we consider identities as elements of $\mathbb{Z}_q^{n}$. In ${\mathbf{SetUp}}$  phase, we choose uniformly random matrix $\bar{\A}$ from $\mathbb{Z}_q^{n \times \bar{m}}$ and a random “short” matrix $\R$ from the Gaussian distribution $D_{\mathbb{Z}, r}^{\bar{m}\times nk }$. 
We construct ${\A} =  \left [ \begin{array}{c  |  r} \bar{\A}  & - \bar{\A}{\R} \end{array} \right ]$ and choose a uniformly random vector ${\u} \in \mathbb{Z}_q^{n}$, where $\R$ is a trapdoor with tag $\bf{0}$. 
We set $({\A},  {\u})$ as the public parameters and  $\R$ as the master secret key. To compute the secret key for an identity $id{_i}$, we construct ${{\A}}_{id{_i}} = \left [ \begin{array}{c  |  r} \bar{\A}  & - \bar{\A}{\R}+ {\H}_{id{_i}}{\G}\end{array} \right ]$, where ${\H}_{id_i}$ is output of FRD\cite{ABB10-EuroCrypt}, $\G$ is the gadget matrix \cite{MP12} and $\R$ is a trapdoor of ${{\A}}_{id{_i}}$ with invertible tag ${\H}_{id_i}$. 
We sample the secret key ${\x}_{id_{i}} \in \mathbb{Z}^{m}$ from $D_{\Lambda_{\u} ^{\perp} ({\A}_{id{_i}} ), s }$, using  $\mathbf{Sample}^{\mathcal{O}}$ with trapdoor $\R$ for ${\A}_{id{_i}}$ and considering $\u$ as syndrome.
The public parameters $({\A},  {\u})$ offers a significant advantage for the simulation: The reduction can embed the LWE instance $\A^{*}$ in the public shared matrix and then sets ${\u} = \left [ \begin{array}{c  |  r} {\A}^{*}  & -{\A}^{*}{\R} \end{array} \right ] \cdot {\x}_{id_{i*}}$, where ${\x}_{id_{i*}}$ is randomly chosen vector from appropriate Gaussian distribution. We treat ${\x}_{id_{i*}}$ as the secret key for target identity $id_{i^*}$ and use it to answer the ReKey query from ${id}^*$. Such clever trick enables collusion resistance for the propose scheme.

We construct ReKey from $id_i$ to $id_j$ as $$rk_{i\rightarrow j} = \left [ \begin{array}{c    c  }{\bf{r}}_1 {\A}_{id_j} & ~~~~{\bf{r}}_1{\u} + {\bf{r}}_2 - {P2}({\x}_{id_i} ) \\ {\bf 0}_{1 \times m}  & {\bf{I}}_{1 \times 1} \end{array} \right ],$$ using secret key of $id_i$.
Apparently, it seems that the structure of ReKey is same as in \cite{SRB14u,SRAS20}. In \cite{SRB14u,SRAS20}, two  matrices, of different order, $\A$ and $\X$ are constructed in public parameters. 
$\A$ is used for encryption and $\X$ is used for re-encryption. In a simplified way, we can say that two different encryption schemes are used to construct $\IBPRE$ of \cite{SRB14u,SRAS20}. 
Similar kind of technique is also used in \cite{NX15}. 
Unfortunately, such technique causes different sizes for original ciphertext and re-encrypted ciphertext. 
Hence, property of proxy transparency is missing in \cite{SRB14u,SRAS20,NX15}. Unlike aforementioned constructions, we use ${\A}_{id{_i}}$ for re-encryption and embed the property of proxy transparency. 
For encryption and decryption, we use dual encryption method from \cite{GPV08}. Since, the order of the original ciphertext and the re-encrypted ciphertext are same in our scheme, we can use same decryption algorithm for both original and re-cncrypted ciphertext. 
But, there are two different decryption algorithms for the original ciphertext and re-encrypted ciphertext in \cite{SRB14u,SRAS20,NX15}. 


For the adaptively-secure $\IBPRE$, we consider identities as ($b_1, b_2, \cdots, b_l) \in\{-1, 1\}^l$. In ${\mathbf{SetUp}}$  phase, we do same as in selectively-secure $\IBPRE$, except choosing one short secret matrix, we choose $l$ random “short” secret matrices ${\R}_1, {\R}_2, \cdots, {\R}_l$ from the same distribution and construct $\bar{\A}_i =  -\bar{\A}{\R}_i $. In contrast of selectively-secure $\IBPRE$, we choose $l + 1$  uniformly random vectors ${\u}_0, {\u}_1, \cdots, {\u}_l$ from $ \mathbb{Z}_q^{n}$. We give $ (\bar{\A}, \bar{\A}_i, {\u}_i)$ as the public parameters and $({\R}_1, {\R}_2, \cdots, {\R}_l)$ as the master secret key. 
To compute the secret key of an identity $id_i =(b_1, b_2, \cdots, b_l)$, we Construct ${\u}_{id_i} = {\u}_0 + \sum_{j=1}^{l} b_j {\u}_j $ and ${\A}_{id_i} =\left [ \begin{array}{c  |  r} \bar{\A}  &  \sum_{j=1}^{l} b_j\bar{\A}_j  + { \G} \end{array} \right ]$, where $\sum_{j=1}^{l} b_j{\R}_j$ is a trapdoor with tag $\bf I$. 
We sample  the secret key ${\x}_{id_{i}}$ from $D_{\Lambda_{{\u}_{id_i}} ^{\perp} ({\A}_{id{_i}} ), s }$ as in Selectively-secure $\IBPRE$. 
For rekey, we do same as in Selectively-secure $\IBPRE$, where $\u$ is replaced by ${\u}_{id{_j}}$. Note that, ${\u}_{id{_j}}$ can be constructed from public parameters and delegatee's identity.

Main challenge in the security reduction of adaptively-secure $\IBPRE$ is that the challenge identity is not known beforehand. It causes {\em abort event} during security reduction. We deal with this issue by using a  family  of  abort-resistant  hash  functions as in \cite{Waters05,ABB10-EuroCrypt}. 
For simulation, we setup public parameters as follows: set $\bar {\A}_i=  -{\A}^{*}{\R}_i + h_i {\G} $, where $h_i$ a secret coefficient from $\mathbb{Z}_q$ for $i =1, \cdots, l$; set ${\u}_0 = {\A}^{*} \cdot {\x}_1$ and ${\u}_i = - {\A}^{*}{\R}_i \cdot {\x}_2$, where $i =1, \cdots, l$ and ${\x}_1, {\x}_2$ are chosen from appropriate Gaussian distribution.
Such setup give leverage to simulate the secret key of challenge identity. We treat  ${\x}^* = \left [ \begin{array}{c } {\x}_1 \\  {\x}_2 \end{array} \right ]$ as the secret key of challenge identity. We use ${\x}^*$ to create challenge ciphertext and to answer ReKey queries from challenge identity in $Phase~2$.
Note that, ${\A}_{id_{i}}
	= \left [ \begin{array}{c  |  r}{ \A}^* &  -{\A}^* \sum_{j=1} b_j{\R}_j +(1 + \sum_{i=1}^{l} h_i b_i)\G\end{array} \right ]$. So, if $(1 + \sum_{i=1}^{l} h_i b_i) = 0$, then the coefficient of $\G$ in ${\A}_{id_{i}}$ is zero. Thus, we cannot sample the secret key of identities for which $(1 + \sum_{i=1}^{l} h_i b_i) = 0$. For these special identities the simulator will be unable to answer key-extraction and ReKey queries, but will be able to construct a useful challenge to solve the given LWE problem instance.
\vspace{-0.6cm}
\section{Preliminaries}
\vspace{-0.3cm}
We denote the real numbers and the integers  by $\mathbb{R}, \mathbb{Z}$, respectively.
We denote column-vectors by lower-case bold letters (e.g. $\bf{b}$), so row-vectors are represented via transposition (e.g. ${\bf{b}}^t$). Matrices are denoted by upper-case bold letters and treat a matrix $ {\X} $ interchangeably with its ordered set $ \{{\bf{x}}_1,{\bf {x}}_2, \ldots\}$ of column vectors. We use ${\bf{I}}$ for the identity matrix and ${\bf{0}}$ for the zero matrix, where the dimension will be clear from context. We use $[ * | * ]$ to denote the concatenation of vectors or matrices.  Singular value \cite{MP12} of ${\B} \in \mathbb{R} ^ {n \times k}$ is denoted by $s_i(\B)$. For ${\x} \in \mathbb{Z}_q^n$, we denote $({\u}_0, \cdots, {\u}_{\lceil \log q \rceil} ) \in \mathbb{Z}_2^{n\cdot {\lceil \log q \rceil} }$ by $BD({\x})$, where $x = \sum _{j=0}^{\lceil \log q \rceil} 2^j \cdot {\u}_j$ and ${\u}_j \in \mathbb{Z}_2^n$. For ${\x} \in \mathbb{Z}_q^n$, we denote $({\x}, 2\cdot {\x}, \cdots, 2^{\lceil \log q \rceil}  \cdot {\x}) \in \mathbb{Z}_2^{n\cdot {\lceil \log q \rceil} }$ by $P2 ({\x})$. By Lemma 2 of \cite{BGV12}, we have $BD({\x})^t \cdot P2({\bf{y})} = {\x}^t  {\bf{y}}$.

A negligible function, denoted generically by $\negl$. We say that a probability is overwhelming if it is $1 - \negl$. The {\em statistical distance} between two distributions ${\bf{X}}$ and ${\bf{ Y}}$ over a countable domain $\Omega$ defined as $\frac 1{2} \sum_{w \in \Omega} | \Pr[{\bf{X}}=w]- \Pr[{\bf{Y}}=w]|.$ We say that a distribution over $\Omega$ is $\epsilon$-far if its statistical distance from the uniform distribution is at most $\epsilon$. Throughout the paper, $r = \omega (\sqrt{\log n})$ represents a fixed function which will be approximated by $\sqrt{ \ln(2n/\epsilon)/\pi}$.

\subsection{Lattices}
\label{1.lattices}
A $lattice ~\Lambda $ is a discrete additive subgroup of $\mathbb{R}^{m}$. Specially, a lattice $\Lambda$ in $\mathbb{R}^{m}$ with basis ${\B}=[{{\b}_1,\cdots,{\b}_n}]\in\mathbb{R}^{m\times n}$, where each ${\b}_i$ is written in column form, is defined as $\Lambda:=\left\{\sum_{i=1}^n{\b}_ix_i | x_i\in\mathbb{Z}~\forall i=1,\ldots,n \right\}\subseteq\mathbb{R}^m.$
We call $n$ the rank of $\L$ and if $n=m$ we say that $\L$ is a full rank lattice. The dual lattice $\Lambda ^{*}$  is the set of all vectors ${ \bf{y}} \in \mathbb{R}^{m} $ satisfying $\langle \bf{x},\bf{y} \rangle \in \mathbb{Z} $  for all vectors ${\bf{x}} \in \Lambda $. If $\B$ is a basis of an arbitrary lattice $\Lambda$, then $ {\B} ^{*} = {\B} ({\B}^t{\B})^{-1}$ is a basis for $\Lambda ^{*}$. For a full-rank lattice, ${\B} ^{*}= {\B}^{-t}$. We refer to $ \tilde{\B}$ as a Gram-Schmidt orthogonalization of $\B$.

In this paper, we mainly consider full rank lattices containing $q\ZZ^m$, called $q$-ary lattices, defined as the following, for a given matrix ${\A}\in\ZZ_q^{n\times m}$ and ${\bf{u}}\in\ZZ_q^n$
	\begin{align*}
	\Lambda ^{\perp} ({\A}) &:= \left\{ {\bf{z}} \in \mathbb{Z}^{m} : {\A} {\bf{z}} =0\!\!\!\mod q \right\}.\\
	\Lambda ({\A }^t) &:=  \left\{ {\bf{z}} \in \mathbb{Z}^{m} : \exists~ {\bf{s}} \in \mathbb{Z}_q^{n}  ~s.t.~ {\bf{z}} = { \A}^t {\bf{s}}\!\!\!\mod q \right\}. \\
	\Lambda _{\bf{u}} ^{\perp} ({\A}) &:= \left\{{\bf{ z}} \in \mathbb{Z}^{m} : {\A} {\bf{z}} ={\bf{u}}\!\!\!\mod q \right\} = \Lambda ^{\perp} ({\A}) + {\bf x} ~for~ {\bf x} \in \Lambda ^{\perp} ({\A}).
	\end{align*}

\noindent Note that, $\Lambda ^{\perp} (\A)$ and $\Lambda ({\A}^t)$ are dual lattices, up to a $q$ scaling factor: $ q \Lambda ^{\perp} ({\A})^{*} = \Lambda ({\A} ^t)$, and vice-versa. Sometimes we consider the non-integral, $1$-$ary$  lattice $ \frac 1{q}\Lambda ({\A} ^t)= \Lambda ^{\perp} ({\A})^{*} \supseteq \mathbb{Z} ^{m}$.

\noindent{\bf Gaussian on Lattices:} Let $\L\subseteq\mathbb{Z}^m$ be a lattice. For a vector ${\bf c}\in{\RR}^m$ and a positive parameter $s\in\RR$, define: $\rho_{{\c}, s}({\x})=\exp\left(\pi\frac{\|{\x}-{\c}\|^2}{s^2}\right)\text{and} ~\rho_{{\c}, s}(\L)=\sum_{\x\in\L}\rho_{{\c}, s}(\x).$
The discrete Gaussian distribution over $\L$ with center $\c$ and parameter $\sigma$ is $D_{{\L},{\c}, s}({\bf{y}})=\frac{\rho_{{\c}, s}(\bf{y})}{\rho_{{\c}, s}(\L)}, \forall {\bf y}\in\L$.

\noindent{\bf Hard Problems on Lattices:} 
\begin{itemize}
\item Consider publicly a prime $q$, a positive integer $n$, and a distribution $\chi$ over $\mathbb{Z}_q$. An $(\mathbb{Z}_q,n,\chi)$-LWE problem instance consists of access to an unspecified challenge oracle $\mathcal{O}$, being either a noisy pseudo-random sampler $\mathcal{O}_{\s}$ associated with a secret ${\s}\in\mathbb{Z}_q^n$, or a truly random sampler $\mathcal{O}_{\$}$ who behaviours are as follows:
		\begin{description}
			\item[$\cal{O}_{\s}$:] samples of the form $({\bf a}_i, v_i)=({\bf a}_i, {\bf a}_i^t {\s} + e_i)\in\mathbb{Z}_q^n\times\mathbb{Z}_q$ where ${\s}\in\mathbb{Z}_q^n$ is a uniform secret key, ${\bf a}_i\in\mathbb{Z}_q^n$ is uniform and $e_i\in\mathbb{Z}_q$ is a noise withdrawn from $\chi$.
			\item[$\cal{O}_\$$:] samples are uniform pairs in $\mathbb{Z}_q^n\times\mathbb{Z}_q$.
		\end{description}\
		The $(\mathbb{Z}_q,n,\chi)$-LWE problem allows responds queries to the challenge oracle $\cal{O}$. We say that an algorithm $\cal{A}$ decides the $(\mathbb{Z}_q,n,\chi)$-LWE problem if 
	 $$\Adv_{\cal{A}}^{\mathsf{LWE}}:=\left|\Pr[{\cal{A}}^{{\cal{O}}_{\s}}= 1] - \Pr[{\cal{A}}^{\cal{O}_{\$}}= 1] \right|$$
  is non-negligible for a random ${\s}\in\mathbb{Z}_q^n$. We denote, $(\mathbb{Z}_q,n,\chi)$-LWE as {LWE}$_{q, \chi}$.
    
Let $\bar{\Psi}_{\alpha}$ be a distribution of the random variable $\lfloor qX\rceil\mod q$, where $\alpha\in(0,1)$ and $X$ is a normal random variable with mean $0$ and standard deviation $\alpha/\sqrt{2\pi}$. It is well known that, under a (quantum) reduction, solving the LWE problem with $\chi = \bar{\Psi}_{\alpha}$ or $D_{\mathbb{Z}, s}$ on average is as hard as the worst case of the approximation version of the shortest independent vector problem, $SIVP_\gamma$, and the decision version of the shortest vector problem, $GapSVP_\gamma$, where $\gamma$ is an approximation factor \cite{Peikert09,Regev09,BLPRS13}.
 
We denote ${\bf{s}}^t {\A} + {\bf{e}}^t \mod q$ for ${\A }\in \mathbb{Z}_q ^{n\times m}$, $ {\bf{s}} \in \mathbb{Z}_q^{n} $ and a Gaussian ${\bf{e}} \in \mathbb{Z}^{m}$ by $g_{\A}({\bf{e}},{\bf{s}})$.
  
 \item The Small Integer Solution (SIS) problem was first suggested to be hard on average by Ajtai \cite{ajtai96} and then formalized by Micciancio and Regev \cite{MR04}. Finding a non-zero short preimage ${\bf{x}}\in \mathbb{Z}^{m}$ such that $f_{\A}({\bf{x}}) = {\A \bf{x}} = {\bf{0}} \mod q $, with $\norm{\bf{x}} \leq \beta$, is an instantiation of the SIS$_{q, n, m, \beta}$ problem.
\end{itemize}

\noindent{\bf Trapdoors for Lattices:} Here, we briefly describe the main results of \cite{MP12}: the definition of $\G$-trapdoor, the algorithms $\bf{Invert^{\mathcal{O}}}$, $\bf{Sample}^{\mathcal{O}}$.
 
 A $\G$-trapdoor is a transformation (represented by a matrix $\R$) from a public matrix $\A$ to a special matrix $\G$ which is called as gadget matrix. The formal definitions as follows:
 
 \begin{defn} [\cite{MP12}]
 Let ${\A }\in \mathbb{Z}_q^{n\times m}$ and ${\G }\in \mathbb{Z}_q^{n\times w}$ be matrices with $m \geq w \geq n$. A $\G$-trapdoor for $\A$ is a matrix ${\R} \in \mathbb{Z}^{(m-w)\times w}$ such that $\A  \left[ \begin{array}{c} \R \\ \bf{I} \end{array}\right ] = \H \G$, for some invertible matrix ${\H} \in \mathbb{Z}_q^{n \times n}$. We refer to $\H$ as the tag or label of the trapdoor. 
 \end{defn} 

 \paragraph{${\bf{Invert}^{\mathcal{O}}} ({\R}, {\A}, {\b}, {\H}_{i})$}\cite{MP12}: On input a vector ${\b}^t = {\s}^t {\A} +{\e}^t$, a matrix\\
 $\A$  and its corresponding $\G$-trapdoor ${\R}$ with invertible tag ${\H}$, the algorithm first computes 
 $${{\b}'}^t = {\b}^t\left[ \begin{array}{c} {\R}  \\ {\bf{I}} \end{array}\right ]$$ and then run the inverting oracle  $\mathcal{O}(\b')$ for $\G$ to get $(\s',\e')$. The algorithm outputs ${\s} = {\H}^{-1} \s' $ and ${\e} = {\b} - {\A}^t {\s}$. Note that, ${\bf{Invert}^{\mathcal{O}}}$ produces correct output if $e_i \in [ - \frac q{4}, \frac q{4})$ i.e. ${\e} \in \mathcal{P}_{1/2}(q \cdot \mathbf{B}^{-t})$, where $\mathbf{B} = {\bf S}_k$ or $\tilde{\bf S}_k$, ${\bf S}_k$ is a basis of $\L^{\perp} ({\G})$; cf. \cite[Section 4.1, Theorem 5.4]{MP12}.

\paragraph{${\bf{Sample}}^{\mathcal{O}} ({\R,\A,{\H}}, {\bf{u}}, s)$} \cite{MP12}: On input  $({\R,\A',\H}, {\bf{u}}, s)$,\\ the algorithm construct $ \A = \left[ \begin{array}{c|c} {\A'} & -{\A}'\R+\H\G \end{array}\right]$, where $\R$ is the $\G$-trapdoor for matrix $\A$ with invertible tag $\H$ and ${\bf{u}} \in \mathbb{Z}_q^{n}$.
The algorithm outputs, using an oracle $\mathcal{O}$ for Gaussian sampling over a desired coset $\Lambda_{\bf{v}} ^{\perp} (\G)$, a vector drawn from a distribution within negligible statistical distance of $D _{\Lambda_{\bf{u}} ^{\perp} ({\A}), s}$.

\subsection{ Identity-Based Unidirectional Proxy Re-Encryption}
\begin{defn} [Identity-Based Unidirectional Proxy ReEncryption (IB-uPRE) \cite{GA07}]
A unidirectional Identity-Based Proxy Re-Encryption $(\IBPRE)$ scheme is a tuple of algorithms  $({\bf{Set Up, Extract, ReKeyGen, Enc, ReEnc, Dec}}):$

\begin{itemize}

\item$(PP,msk) \longleftarrow {\bf{SetUp}}(1^n):$ On input the security parameter $1^{n}$, the $\bf{set up}$ algorithm outputs $PP, msk$.

\item $ sk_{id} \longleftarrow {\bf{Extract}}(PP, msk, id):$ On input an identity $id$, public parameter PP, master secret key msk, output the secret key $sk_{id}$ for $id$.

\item $rk _{i \rightarrow j} \longleftarrow\mathbf{ReKeyGen}(PP, sk_{id_{i}}, {id_i}, {id_j}):$ On input a public parameter $PP$, secret key $sk_{id_{i}}$ of a delegator $i$, and the identities of delegator $i$ and delegatee $j$, $id_i, id_j$ respectively, output a unidirectional re-encryption key $rk _{i \rightarrow j} $.

\item $ct \longleftarrow {\bf{Enc}}(PP, id, m):$ On input an identity $id$, public parameter $PP$ and a plaintext $m \in \mathcal{M}$, output a ciphertext $ct$ under the specified identity $id$.

\item $ct' \longleftarrow {\bf{ReEnc}}(PP,rk _{i \rightarrow j}, ct):$ On input a ciphertext $ct$ under the identity $i$  and a re-encryption key $rk _{i \rightarrow j}$, output a ciphertext $ct'$ under the identity $j$.

\item $m \longleftarrow {\bf{Dec}}(PP,sk_{id_{i}}, ct):$ On input the ciphertext $ct$ under the identity $i$ and secret key $sk_{id_{i}}$ of $i$, the algorithm outputs a plaintext $m$ or the error symbol $\bot$.

\end{itemize} 
\end{defn}

An Identity-Based Proxy Re-Encryption scheme is called  single-hop if a ciphertext can be re-encrypted only once. In a multi-hop setting   proxy can apply further re-encryptions to already re-encrypted ciphertext. 

\begin{defn} [Single-hop IB-uPRE Correctness]
A single-hop unidirectional Identity-Based Proxy Re-Encryption scheme $({\bf{Set Up, Extract, ReKeyGen, Enc}},$ ${\bf{ReEnc, Dec}})$ decrypts correctly for the message $m \in \mathcal{M}$ if $:$
\begin{itemize}
\item For all $ sk_{id}$ output by ${\bf{Extract}}$ under $id$ and for the message $m \in \mathcal{M}$,\\ it holds that $ {\bf{Dec}}(PP, sk_{id}, {\bf{Enc}}(PP, id, m))  = m$.

\item For any re-encryption key $rk _{i \rightarrow j}$ output by $\mathbf{ReKeyGen} (PP, sk_{id_{i}}, {id_i}, {id_j})$ and any $ct={\bf{Enc}}(PP, id_i, m)$ it holds that\\  ${\bf{Dec}}(PP, sk_{id_{j}}, {\bf{ReEnc}}(PP,rk _{i \rightarrow j}, ct)) = m$.

\end{itemize}
\end{defn}

\paragraph{{\bf Security Game of Selectively Secure Identity-Based Unidirectional Proxy \\Re-Encryption Scheme against Chosen Plaintext Attack (IND-sID-CPA)}}
To describe the security model we first classify all of the users into honest $(HU)$ and corrupted $(CU)$. In the honest case an adversary does not know secret key, whereas for a corrupted user the adversary has secret key.
Let $\mathcal{A}$ be the PPT adversary and $\Pi = ({\bf{Set Up, Extract, ReKeyGen}},$ ${\bf{Enc, ReEnc, Dec}})$ be an $\IBPRE$ scheme with a plaintext space $\mathcal{M}$ and a ciphertext space $\mathcal{C}$. Let $id^{*}(\in HU)$ be the target user. Security game is defined according to the following game $\Exp_{\mathcal{A}}^ {\INDsCPA} (1^{n}):$

\begin{enumerate}
\item $\bf{Set Up}$: The challenger runs ${\bf{Set Up}}(1^{n})$ to get ($PP, msk)$ and give $PP$ to $\mathcal{A}$. 
\item{{\bf Phase 1:}} The adversary $\mathcal{A}$ may make queries polynomially many times in any order to the following oracles:
\begin{itemize}
\item $\mathcal{O}^{\bf{Extract}}$: an oracle that on input $id \in CU$, output $sk_{id}$; Otherwise, output $\bot$.

\item $\mathcal{O}^{\bf{ReKeyGen}}$: an oracle that on input the identities of $i$-th and $j$-th users: if $id_i \in HU $, $id_j \in HU$ or $id_i, id_j \in CU$ or $id_i \in CU, id_j \in HU$, output $rk _{i \rightarrow j}$; otherwise, output $\bot$.

\item $\mathcal{O}^{\bf{ReEnc}}$: an oracle that on input the identities of $i, j$-th users and ciphertext of $i$-th user: if $id_i, id_j \in HU$ or $id_i, id_j \in CU$ or $id_i \in CU, id_j \in HU $output re-encrypted ciphertext; otherwise, output $\bot$.
\end{itemize}

\item $\bf{Challenge}$: $\mathcal{A}$ outputs a message $m \in \mathcal{M}$. The challenger picks a random bit $r \in \{0, 1\}$ and a random ciphertext $C$ from the ciphertext space. If $ r=0$ it sets the challenge ciphertext $ct^* = Enc(PP, id^*, m)$. If $r=1$ it sets the challenge ciphertext $ct^* = C$. Then challenger sends the $ct^*$ as the challenge to the adversary.

\item{{\bf Phase 2:}} After receiving the challenge ciphertext, $\mathcal{A}$ continues to have access to the $\mathcal{O}^{\bf{Extract}}$, $\mathcal{O}^{\bf{ReKeyGen}}$ and $\mathcal{O}^{\bf{ReEnc}}$ oracle as in {\bf Phase 1}.

\item $\mathcal{O}^{\bf{Decision}}$: On input $r'$ from adversary  $\mathcal{A}$, this oracle outputs $1$ if $r = r'$ and $0$ otherwise. 
\end{enumerate}

\noindent The advantage of an adversary in the above experiment $\Exp_{\mathcal{A}}^ {\INDsCPA}$$(1^{n})$ is defined as $|\Pr[r' = r]-\frac1{2} |$.

\begin{defn}
\label{def:security}
An $\IBPRE$ scheme is $\INDsCPA$ secure if all PPT adversaries $\mathcal{A}$ have at most a negligible advantage in experiment  $\Exp_{\mathcal{A}}^ {\INDsCPA}(1^{n})$.
\end{defn}

For the Adaptive-Identity, instead of announcing the challenge identity at the starting of the game, Adversary will announce it at the time of challenge phase. Only constraints is that, there was no Extract queries on that challenge identity before. The resulting security notion is defined using the modified game as in Definition \ref{def:security}, and is denoted {\bf IND–ID-CPA}.

\section{Selectively Secure Identity-Based Unidirectional Proxy Re-Encryption Scheme ($\SIBPRE$)}
\subsection{Construction of Selective-IB-uPRE}\label{sec:sel_construction}
In this section, we present our construction of $\SIBPRE$. In the following construction, we encode identities as follows:  
\begin{itemize}
\item{\bf Encoding of Identity:} To encode identity, we use {\em full-rank difference} map (FRD) as in \cite{ABB10-EuroCrypt}. FRD: $\mathbb{Z}_q^{n} \rightarrow \mathbb{Z}_q^{n\times n}$; $id \mapsto {\H}_{id}$. We assume identities are non-zero elements in $ \mathbb{Z}_q^{n} $. The set of identities can be expanded to $\{0,1\}^{*}$ by hashing identities into $ \mathbb{Z}_q^{n} $ using a collision resistant hash. FRD satisfies the following properties:
\begin{enumerate}
\item $\forall ~ distinct ~id_1, id_2 \in  \mathbb{Z}_q^{n}$, the matrix ${\H}_{id_1} - {\H}_{id_2} \in \mathbb{Z}_q^{n\times n}$ is full rank; 
\item $\forall ~ id \in  \mathbb{Z}_q^{n} \setminus \{{\bf 0}\}$, the matrix ${\H}_{id} \in \mathbb{Z}_q^{n\times n}$ is full rank; 
\item FRD is computable in polynomial time (in $n \log q$).
\end{enumerate}
\end{itemize}

We set the parameters as the following.
\begin{itemize}
\item $\mathbf{G}\in\mathbb{Z}_q^{n \times nk}$ is a gadget matrix for large enough prime power $ q = p^ {e}  = poly (n)$ and $k = {\lceil \log q \rceil} = O (\log n) $, so there are efficient algorithms to invert $ g_\mathbf{G} $ and to sample for $ f_\mathbf{G}$. 

\item $\bar{m} = O(nk) $ and  the Gaussian $ \mathcal{D} = D_{\mathbb{Z}, r}^{\bar{m}\times nk }$, so that $( \bar{\A}, \bar{\A} \R)$ is negl(n)-far from uniform for $\bar{\A}$.

\item The LWE error rate $\alpha $  for $\SIBPRE$ should satisfy $1 / \alpha = O (nk)^{2} \cdot r^2$. 

\end{itemize}

The proposed $\SIBPRE$ consists of the following algorithms: 

\paragraph{$\mathbf{SetUp}(1^n)$} On input a security parameter $n$, do:
\begin{enumerate}
\item Choose $\bar{\A} \leftarrow\mathbb{Z}_q^{n \times \bar{m}} $, $\R \leftarrow \mathcal{D}$, and construct ${\A} =  \left [ \begin{array}{c  |  r} \bar{\A}  & - \bar{\A}{\R} \end{array} \right ]$ $\in \mathbb{Z}_q^{n \times m}$, where $ m = \bar{m} + nk $. 

\item Choose a uniformly random vector ${\u}$ from $ \mathbb{Z}_q^{n}$.

\item Output the public parameter $PP= ({\A},  {\u})$ and the master secret key is $msk = \R$. \end{enumerate}

\paragraph{$\mathbf{Extract}( PP, msk, id)$} On input a public parameter $PP$, master secret key $msk$ and the identity of $i$-th user $id_i$, do:
\begin{enumerate}
\item Construct ${{\A}}_{id{_i}} = \left [ \begin{array}{c  |  r} \bar{\A}  & - \bar{\A}{\R}+$${\H}_{id{_i}}$$\G$$\end{array} \right ] \in\mathbb{Z}_q^{n \times m}$. 

\item Sample ${\x}_{id_{i}} \in \mathbb{Z}^{m}$ from $D_{\Lambda_{\u} ^{\perp} ({\A}_{id{_i}} ), s }$, using  $\mathbf{Sample}^{\mathcal{O}}$ with trapdoor $\R$ for ${\A}_{id{_i}}$. 
\item Output the secret key as $sk_{id_{i}}=  {\x}_{id_{i}} \in \mathbb{Z}^{m}$. 

\end{enumerate}

\paragraph{$\mathbf{Enc}(PP, id_{i}, b)$} On input a public parameter $PP$, the identity of $i$-th user $id_i$ and a message $b \in \{0,1\}$, do:
\begin{enumerate} 
\item Construct ${{\A}}_{id{_i}} = \left [ \begin{array}{c  |  r} \bar{\A}  & - \bar{\A}{\R}+$${\H}_{id{_i}}$$\G$$\end{array} \right ]  $ $\in\mathbb{Z}_q^{n \times m}$. 
 \item Choose a uniformly random  ${\bf s} \leftarrow \mathbb{Z}_q^{n}$.
\item Sample error vectors $e \leftarrow  D_{\mathbb{Z},\alpha q}, {\e}_{0} \leftarrow  D_{\mathbb{Z},\alpha q}^{\bar{m}}$   and    ${\e}_{1}\leftarrow D_{\mathbb{Z},s'}^{nk}$, where $s'^{2} = (\| {\e}_{0} \| ^ {2} + \bar {m} (\alpha q)^{2}) r^2$. Let the error vector $ {\e} = ({\e}_{0},  {\e}_{1}) \in \mathbb{Z} ^{m}$.
\item Compute $ {\c}_1 = {{\A}}_{id{_i}}^t {\bf s} + {\e} \mod q \in \mathbb{Z}_{q}^{m}$ and  $ {\c}_2 = {\u}^t {\bf s} + e +b\cdot \lfloor q/2 \rfloor  \mod q \in \mathbb{Z}_{q}$.
\item Output the ciphertext $ct = ({\c}_1, {\c}_2) \in\mathbb{Z}_{q}^{m} \times \mathbb{Z}_{q}$.
\end{enumerate}

\paragraph{$\mathbf{Dec}(PP, sk_{id_{i}}, ct)$} On input a public parameter $PP$, the secret key of $i$-th user $sk_{id_{i}}$ and ciphertext $ct$, do:
\begin{enumerate}
\item Compute $b' = {\c}_2 - {\x}_{id_{i}}^t {\c}_1 \in \mathbb{Z}_{q}$.
\item Output $0$ if $b'$ is closer to $0$ than to $\lfloor q/2 \rfloor \mod q$; Otherwise output $1$.
\end{enumerate}

\paragraph{$\mathbf{ReKeyGen}(PP, sk_{id_{i}}, id_i, id_j)$} On input a public parameter $PP$, the secret key of $i$-th user $sk_{id_{i}}$ and identity of $j$-th user $id_j$, do:
\begin{enumerate}
\item Construct ${\A}_{id_i}$ and ${\A}_{id_j}$.
\item Choose ${\bf{r}}_{1}\leftarrow  D_{\mathbb{Z}, r}^{{mk}\times n }$ and ${\bf{r}}_{2}\leftarrow D_{\mathbb{Z}, r}^{{mk}\times 1 } $.
\item Construct the proxy re-encryption key  \\ $rk_{i\rightarrow j} = \left [ \begin{array}{c    c  }{\bf{r}}_1 {\A}_{id_j} & ~~~~{\bf{r}}_1{\u} + {\bf{r}}_2 - {P2}({\x}_{id_i} ) \\ {\bf 0}_{1 \times m}  & {\bf{I}}_{1 \times 1} \end{array} \right ]  \in \mathbb{Z} _{q}^{{(mk+1)}\times{(m+1)}} $
\item Output $rk_{i\rightarrow j}$.
\end{enumerate}

\paragraph{$\mathbf{ReEnc}( rk_{i\rightarrow j}, ct)$} On input $rk_{i\rightarrow j}$ and $i$-th user's ciphertext $ct = ({\c}_1, {\c}_2)$, Compute:
\begin{enumerate}
\item Compute the re-encrypted ciphertext $\bar{ct} = (\bar{\c}_1, \bar{\c}_2)$ as follows: \\$\bar{ct}^t = \left [ \begin{array}{c|c} BD({\c}_1)^t & {\c}_2^t \end{array} \right] \cdot rk_{i\rightarrow j} \in \mathbb{Z}_q ^{1 \times (m+1)}$
\item Output the re-encrypted ciphertext $ \bar{ct}$.
\end{enumerate}

 

\subsection{Correctness and Security}
In this section, we analyze the correctness and security of the proposed scheme. In respect of correctness, the main point is to consider the growth of error due to re-encryption. We have proved that the growth of error is controlled in Theorem \ref{sel_correctness}. Further, we have proved security of the construction in the selective identity based model, according to the Definition \ref{def:security}, against chosen plaintext attack in Theorem \ref{thm:sel_security}


\begin{thm} [Correctness] 
\label{sel_correctness}
The $\SIBPRE$ scheme with parameters proposed in Section~\ref{sec:sel_construction}  is correct.
\end{thm}

\noindent{\bf Proof:}
To show that the decryption algorithm outputs a correct plaintext, we will consider both original and re-encrypted ciphertext.  Let 
$sk_{id_{i}}= {\x}_{id_{i}}$ and 
$sk_{id_{j}}={\x}_{id_{j}}$ 
be the secret key for $i$-th and $j$-th user respectively.  \\
From $\mathbf{ReKeyGen}(PP, sk_{id_{i}}, id_i, id_{j})$ algorithm, we get 
$$ rk_{i\rightarrow j}= \left [ \begin{array}{c    c  }{\bf{r}}_1 {\A}_{id_j} & ~~~~{\bf{r}}_1{\u} + {\bf{r}}_2 - {P2}({\x}_{id_i} ) \\ {\bf 0}_{1 \times m}  & {\bf{I}}_{1 \times 1} \end{array} \right ].$$

Let  $ ct= ({\c}_1, {\c}_2)$ be the ciphertext of a message ${b} \in \{0,1\}$ for $i$-th user and $\bar{ct} = (\bar{\c}_1, \bar{\c}_2)=(\mathbf{ReEnc}( PP, rk_{i \rightarrow j}, ct ))$  be the re-encrypted ciphertext for the $j$-th user. Thus, we need to prove that $\mathbf{Dec} (PP, sk_{id_{i}}, ct) = \mathbf{Dec} (PP, sk_{id_{j}}, ct') = {b}$.

First we decrypt the original ciphertext, 
\begin{align*} 
 b' &= {\c}_2 - {\x}_{id_{i}}^t {\c}_1\\
      &= {\u}^t {\bf s} + e +b\cdot \lfloor q/2 \rfloor - {\x}_{id_{i}}^t ({{\A}}_{id{_i}}^t {\bf s} + {\e})\\
      &= {\u}^t {\bf s} + e +b\cdot \lfloor q/2 \rfloor - {\u}^t {\bf s} - {\x}_{id_{i}}^t  {\e}\\
      &= \underbrace{e - {\x}_{id_{i}}^t  {\e}}_{\text{ error }} +b\cdot \lfloor q/2 \rfloor 
 \end{align*}

To get a correct decryption, the norm of the error term should be less than $q/4$ i.e. $|  e - {\x}_{id_{i}}^t  {\e}| < q/4$.
Let us estimate the norm of noises,
we have $s_1({\R}) \leq O (\sqrt {nk}) \cdot r$ \cite[ Lemma 2.9]{MP12}, $s_1({\x}_{id_{i}})\leq 2\sqrt {6}  \cdot \sqrt{ m} \cdot \sqrt {s_1({\R})^{2} +1} \cdot r$. We have ${\e} = (\bar{\e}_{0}, {\e}_{1})$. By \cite[Lemma 12]{ABB10-EuroCrypt}, \cite[Lemma 1.5]{ba1993}, $\|{ \e}_{0}\| < \alpha q \sqrt{\bar{m}}$ and $\|{\e}_{1}\| < \alpha q \sqrt{2\bar{m}nk} \cdot r$ i.e. $\|{\e}\| < 2\alpha q \sqrt{2\bar{m}nk} \cdot r$. So, $|  e - {\x}_{id_{i}}^t  {\e}| < \alpha q \cdot O (nk)^2 \cdot r^2$. As $1 / \alpha = O (nk)^{2} \cdot r^2$, we have $|  e - {\x}_{id_{i}}^t  {\e}| < \alpha q \cdot O (nk)^2 \cdot r^2 < q/4$.

For the re-encrypted ciphertext $ \bar{ct}$ for $id_j$, we have $\bar{ct}^t = \left [ \begin{array}{c|c} BD({\c}_1)^t & {\c}_2 \end{array} \right] \cdot rk_{i\rightarrow j}$. We have, 
\begin{align*}
 {b'}^t &= \bar{\c}^t_2 - \bar{\c}^t_1 {\x}_{id_{j}}\\
         &= \bar{ct}^t \cdot \left [ \begin{array}{c} -{\x}_{id_j}\\ 1 \end{array} \right]\\
         &= \left [ \begin{array}{c|c} BD({\c}_1)^t & {\c}_2^t \end{array} \right] \cdot rk_{i\rightarrow j} \cdot \left [ \begin{array}{c} -{\x}_{id_j}\\ 1 \end{array} \right]\\
         &= \left [ \begin{array}{c|c} BD({\c}_1)^t & {\c}_2^t \end{array} \right] \cdot \left [ \begin{array}{c    c  } {\bf{r}}_1 {\A}_{id_j} & ~~~~{\bf{r}}_1{\u} + {\bf{r}}_2 - {P2}({\x}_{id_i} ) \\ {\bf 0}_{1 \times m}  & {\bf{I}}_{1 \times 1} \end{array} \right ] \cdot \left [ \begin{array}{c} -{\x}_{id_j}\\ 1 \end{array} \right]\\
         &=\left [ \begin{array}{c|c} BD({\c}_1)^t & {\c}_2^t \end{array} \right] \cdot \left [ \begin{array}{c}- {\bf{r}}_1 {\A}_{id_j} {\x}_{id_i} + {\bf{r}}_1{\u} + {\bf{r}}_2 - {P2}({\x}_{id_i}) \\ 1\end{array} \right ]\\
         &=\left [ \begin{array}{c|c} BD({\c}_1)^t & {\c}_2^t \end{array} \right] \cdot \left [ \begin{array}{c}  {\bf{r}}_2 - {P2}({\x}_{id_i}) \\ 1\end{array} \right ]\\
         &= BD({\c}_1)^t {\bf{r}}_2- BD({\c}_1)^t {P2}({\x}_{id_i})+ {\c}_2^t\\
         &=BD({\c}_1)^t {\bf{r}}_2- {\c}_1^t {\x}_{id_i}+ {\c}_2^t\\
    {b'} &= {\bf{r}}_2^t BD({\c}_1)-{\x}_{id_i}^t {\c}_1 + {\c}_2 \\
         &={\bf{r}}_2^t BD({\c}_1)-{\x}_{id_i}^t({{\A}}_{id{_i}}^t {\bf s} + {\e}) +  {\u}^t {\bf s} + e +b\cdot \lfloor q/2 \rfloor\\
         &= \underbrace{{\bf{r}}_2^t BD({\c}_1) - {\x}_{id_i}^t {\e} + e }_{\text{error}}+b\cdot \lfloor q/2 \rfloor
\end{align*}

To get a correct decryption, the norm of the error term should be less than $q/4$ i.e. $|  {\bf{r}}_2^t BD({\c}_1) - {\x}_{id_i}^t {\e} + e | < q/4$. Let us estimate the norm of noises, we have  $s_1({\bf{r}}_2) \leq O (\sqrt {nk}) \cdot r$ \cite[ Lemma 2.9]{MP12} , 
 $|{\bf{r}}_2^t BD({\c}_1)| \leq O (nk)^{1.5} \cdot r$. Also, $|  e - {\x}_{id_{i}}^t  {\e}| < \alpha q \cdot O (nk)^2 \cdot r^2$.
As $1 / \alpha = O (nk)^{2} \cdot r^2$, $| {\bf{r}}_2^t BD({\c}_1) - {\x}_{id_i}^t {\e} + e | < \alpha q \cdot O (nk)^2 \cdot r^2 < q/4$.
\qed

\begin{thm}[Security]
\label{thm:sel_security}
The above scheme is $\INDsCPA$ secure assuming the hardness of decision-{\em LWE}$_{q, \chi}.$ 
\end{thm}
\noindent{\bf Proof:}
Let the LWE samples of the form $({\bf a}_i, v_i) = ({\bf a}_i, {\bf a}_i^t {\s}+ e_i) \in \mathbb{Z}_q^{n} \times \mathbb{Z}_{q}$, where ${\bf s} \leftarrow \mathbb{Z}_q^{n}$, uniformly random and $e_i \in \mathbb{Z}_{q} $, sample from $\chi$, ${\bf a}_i$ is uniform in $\mathbb{Z}_q^{n}$. we construct column-wise matrix $\A^{*}$ from these samples ${\bf a}_i$ and a vector ${\v}^{*}$ from the corresponding $v_i$. Let $id_{i^*}$ be the target user. The proof proceeds in a sequence of games.

\noindent{\bf Game 0:} This is the original $\INDsCPA$ game from definition between an adversary $\mathcal{A}$ against scheme and an $\INDsCPA$ challenger.

\noindent{\bf Game 1:} In ${\bf Game~1}$ we change the way that the challenger generates ${\A}, {\u}$ in the public parameters. In $\bf{SetUp}$ phase, do as follows:
	\begin{itemize}
	 \item Set the public parameter $ \bar {\A} = \A^{*}$, where $\A^{*}$ is from LWE instance $(\A^{*}, {\v}^*)$ and set ${\A} =  \left [ \begin{array}{c  |  r} {\A}^{*}  & -{\A}^{*}{\R} - {\H}_{id_{i^*}}\G \end{array} \right ]$, where $\R$ is chosen in the same way as in ${\bf Game~0}$. 
	 \item Choose ${\x}_{id_{i*}}\leftarrow \mathcal{D}= D_{\mathbb{Z}, r}^{m \times 1}$; Set ${\u} = \left [ \begin{array}{c  |  r} {\A}^{*}  & -{\A}^{*}{\R} \end{array} \right ] \cdot {\x}_{id_{i*}}$.
\item Set $PP = ({\A}, {\u})$ and send it to the Adversary $\mathcal{A}$.
	
	\item $\mathcal{O}^{\bf{Extract}}$: To answer a secret key query against $id_i\in CU$, challenger will do the following: Construct	${\A}_{id_{i}} =
	\left [ \begin{array}{c|r} 
	{\A}^*  &  -{\A}^* {\R} -{\H}_{id_{i^*}}{\G}+ {\H}_{id{_i}}{\G} \end{array} \right ]  
	= \left [ \begin{array}{c  |  r}{ \A}^* &  -{\A}^* {\R} +({\H}_{id{_i}}-{\H}_{id_{i^*}})\G
\end{array} \right ].$
So, $\R$ is a trapdoor of ${\tilde \A_{i}}$ with invertible tag $({\H}_{id{_i}}-{\H}_{id_{i^*}})$. Then from $\mathbf{Extract}$ algorithm, challenger gets the secret key $sk_{id_{i}}= {\x}_{id_{i}}$ for $id_i$, sends $sk_{id_{i}}$ to the adversary $\mathcal{A}$.\\ 
Challenger will send $\bot$, against the secret key query for $id_i\in HU$.


\item $\mathcal{O}^{\bf{ReKeyGen}}$: For the re-encryption key query from ${id_{i^*}}$ to $id_j\in HU$, challenger will compute ${\A}_{id_j}$, then $$rk_{i^*\rightarrow j} = \left [ \begin{array}{c    c  } {\bf{r}}_1 {\A}_{id_j} & {\bf{r}}_1{\u} + {\bf{r}}_2 - {P2}({\x}_{id_{i^*}} ) \\ {\bf 0}_{1 \times m}  & {\bf{I}}_{1 \times 1} \end{array} \right ], $$ where ${\bf{r}}_1, {\bf{r}}_2$ was chosen as in the scheme.
 For other re-encryption key query challenger maintain the restrictions as in definition \ref{def:security} and computes $rk_{i \rightarrow j}$ according to the $\mathbf{ReKeyGen}$ algorithm to reply the adversary. 
 \item $\mathcal{O}^{\bf{ReEnc}}$: For re-encryption query challenger maintain the restrictions as in definition \ref{def:security} and computes $\mathbf{ReEnc}(rk_{i \rightarrow j}, ct)$ according to the $\mathbf{ReEnc}$ algorithm to reply the adversary. 

Due to left-over hash lemma \cite[Lemma 14]{ABB10-EuroCrypt}, $({\A}^{*}, -{\A}^{*}{\R}, \left [ \begin{array}{c  |  r} {\A}^{*}  & -{\A}^{*}{\R} \end{array} \right ] \cdot {\x}_{id_{i*}})$ is statistically indistinguishable with uniform distribution.\\ Hence, $({\A}^{*}, -{\A}^{*}{\R}- {\H}_{id_{i^*}}{\G},  \left [ \begin{array}{c  |  r} {\A}^{*}  & -{\A}^{*}{\R} \end{array} \right ] \cdot {\x}_{id_{i*}})$ is statistically indistinguishable with uniform distribution. Furthermore, ${\A}, {\u}$ and responses to key queries are statistically close to those in ${\bf Game~0}$. Hence, ${\bf Game ~0}$ and ${\bf Game ~1}$ are statistically indistinguishable. 
\end{itemize}

\noindent{\bf Game 2:} In ${\bf Game~2}$, we change the way that the challenger generates challenge ciphertext. Here Challenger produces the challenge ciphertext ${ct}^* = ({\c}_1^*, {\c}_2^*)$ for the target identity $id_{i^*}$ on a message  ${{b}^*}\in \{0,1\}$ as follows: 
\begin{itemize} 
\item Choose a uniformly random  ${\bf s} \leftarrow \mathbb{Z}_q^{n}$ and noise vectors $  {\e}_{0} \leftarrow  D_{\mathbb{Z},\alpha q}^{\bar{m}}$.

  Set $ {\c}_1^* =  \left [ \begin{array}{c}{\v}^{*} \\ -{\R}^t {\v}^{*} \end{array} \right ]  \in \mathbb{Z}_{q}^{m}$  and $ {\c}_2^* =  {\x}_{id_{i^*}}^t  \cdot \left [ \begin{array}{c} {\v}^{*} \\ -{\R}^t {\v}^{*} \end{array} \right ] +b^{*} \cdot \lfloor q/2 \rfloor  \in \mathbb{Z}_{q}$, where ${\v}^{*} = {{\A}^{*}}^t \cdot{\s} + {\e}_0$.  [ $({\A}^{*}, {\v}^{*})$ be the LWE instance]
 
  Send $ct^* = ({\c}_1^*, {\c}_2^*)$ to the adversary.
 \item Here ${\c}_1^*, {\c}_2^*$ satisfies,  
  \begin{align*}
	{\c}_1^*  &=  \left [ \begin{array}{c}{\v}^{*} \\ -{\R}^t {\v}^{*} \end{array} \right ]\\
	              &=   \left [ \begin{array}{c} {\A}^{*t} \cdot{\s} + {\e}_0 \\ -{\R}^t {\A}^{*t} \cdot{\s} - {\R}^t {\e}_0 \end{array}\right ] \\
	               &=  \left [ \begin{array}{c} {\A}^{*t}  \\ -{\R}^t {\A}^{*t} \end{array} \right ] \cdot{\s} +  \left [ \begin{array}{c} {\e}_0 \\ - {\R}^t {\e}_0  \end{array} \right ] \\
	              &=  {{\A}}_{id{_{i^*}}}^t {\bf s} +  \left [ \begin{array}{c} {\e}_0 \\ - {\R}^t {\e}_0 \end{array} \right ]  \in \mathbb{Z}_{q}^{m}
	\end{align*}

 \begin{align*}
	{\c}_2^*  &= {\x}_{id_{i^*}}^t  \cdot \left [ \begin{array}{c} {\v}^{*} \\ -{\R}^t {\v}^{*} \end{array} \right ] +b^{*} \cdot \lfloor q/2 \rfloor \\
	&= {\x}_{id_{i^*}}^t  \cdot \left [ \begin{array}{c} {\A}^{*t} \cdot{\s} + {\e}_0 \\ -{\R}^t {\A}^{*t} \cdot{\s} - {\R}^t {\e}_0 \end{array}\right ] +b^{*} \cdot \lfloor q/2 \rfloor\\
        &= {\x}_{id_{i^*}}^t  \cdot {{\A}}_{id{_{i^*}}}^t {\bf s} +  {\x}_{id_{i^*}}^t  \left [ \begin{array}{c} {\e}_0 \\ - {\R}^t {\e}_0  \end{array} \right ] +b^{*} \cdot \lfloor q/2 \rfloor\\
	&= {\u}^t {\s} + {\x}_{id_{i^*}}^t  \left [ \begin{array}{c} {\e}_0 \\ - {\R}^t {\e}_0  \end{array} \right ] +b^{*} \cdot \lfloor q/2 \rfloor \in \mathbb{Z}_{q}
\end{align*}
By Corollary 3.10 in \cite{Regev05}, the noise term $( {\e}_0, - {\R}^t {\e}_0)$ of ${\c}^*_1 $ is within $ \negl$ statistical distance from discrete Gaussian distribution $D_{\mathbb{Z},s'}^{nk}$. The same argument, also, applies for the noise term of ${\c}^*_2$. So, $({\c}_1^*, {\c}_2^*)$ is the valid challenge ciphertext in ${\bf Game~2}$\cite[Lemma 12, Lemma 14]{ABB10-EuroCrypt}. 
\end{itemize}
Hence, ${\bf Game ~1}$ and ${\bf Game ~2}$ are statistically indistinguishable.

\noindent{\bf Game 3:} Here, we only change how the ${\v}^{*}$ component of the challenge ciphertext is created, letting it be uniformly random in $\mathbb{Z}_{2q} ^{\bar m}$. Challenger constructs the public parameters, answers the secret key queries, re-encryption queries and construct the remaining part of the challenge ciphertext exactly as in Game~2. It follows from the hardness of the decisional LWE$_{q, \chi}$ that ${\bf Game ~2}$ and ${\bf Game~3}$ are computationally indistinguishable.

Now, by the left-over hash lemma \cite[Lemma 14]{ABB10-EuroCrypt}, (${ \A}^*,{\v}^{*}, -{ \A}^{*}{\R},-{\R}^t{{\v}^{*}}$) is $\negl$-uniform when ${ \R}$ is chosen as in Game 2. Therefore, the challenge ciphertext has the same distribution (up to  $\negl$ statistical distance) for any encrypted message. So, the advantage of the adversary against the proposed scheme is same as the advantage of the attacker against decisional LWE$_{q, \chi}$. This completes the proof.
\qed

\section{Adaptively Secure Identity-Based Unidirectional Proxy Re-Encryption Scheme ($\AIBPRE$)}
\subsection{Construction of Adaptive-IB-uPRE}\label{sec:adap_construction}
In this section, we present our construction of $\AIBPRE$. We set the parameters as in section \ref{sec:sel_construction}. The proposed $\AIBPRE$ consists of the following algorithms:

\paragraph{$\mathbf{SetUp}(1^n)$}
\begin{enumerate}
\item Choose $\bar{\A} \leftarrow\mathbb{Z}_q^{n \times \bar{m}} $ and ${\R}_1, {\R}_2, \cdots, {\R}_l \leftarrow \mathcal{D}= D_{\mathbb{Z}, r}^{\bar{m}\times nk }$, Construct $\bar{\A}_i =  -\bar{\A}{\R}_i $ $\in \mathbb{Z}_q^{n \times nk}$ for $i = 1,\cdots, l$. 
\item Choose $l + 1$  uniformly random vectors ${\u}_0, {\u}_1, \cdots, {\u}_l$ from $ \mathbb{Z}_q^{n}$.

\item Output the public parameter $PP= (\bar{\A}, \bar{\A}_1, \bar{\A}_2, \cdots, \bar{\A}_l, {\u}_0, {\u}_1, \cdots, {\u}_l)$ and the master secret key is $msk =({\R}_1, {\R}_2, \cdots, {\R}_l)$.

\end{enumerate}

\paragraph{$\mathbf{Extract}( PP, MK, id)$} On input public parameter PP, the master key MK and an identity of $i$-th user, $id_i$ = ($b_1, b_2, \cdots, b_l$) $\in\{-1, 1\}^l$:
\begin{enumerate}
\item Construct ${\A}_{id_i} =\left [ \begin{array}{c  |  r} \bar{\A}  &  \sum_{j=1}^{l} b_j\bar{\A}_j  + { \G} \end{array} \right ] \in\mathbb{Z}_q^{n \times m}$, where $m = \bar{m} + nk$. Here, $\sum_{j=1}^{l} b_j{\R}_j$ is a trapdoor of ${\A}_{id_i}$ with tag $\bf I$.
\item Construct ${\u}_{id_i} = {\u}_0 + \sum_{j=1}^{l} b_j {\u}_j $.
\item Sample ${\x}_{id_{i}} \in \mathbb{Z}^{m}$ from $D_{\Lambda_{{\u}_{id_i}} ^{\perp} ({\A}_{id{_i}} ), s }$, using  $\mathbf{Sample}^{\mathcal{O}}$ with trapdoor $\sum_{j=1}^{l} b_j{\R}_j$ for ${\A}_{id{_i}}$.
\item Output the secret key $sk_{id_{i}}=  {\x}_{id_{i}} \in \mathbb{Z}^{m}$. 
\end{enumerate}

\paragraph{$\mathbf{Enc}(PP, id_{i}, b)$} On input a public parameter $PP$, the identity of $i$-th user $id_i$ and a message $b \in \{0,1\}$, do:
\begin{enumerate} 
\item  Construct ${\A}_{id_i} =\left [ \begin{array}{c  |  r} \bar{\A}  &  \sum_{j=1}^{l} b_j\bar{\A}_j  + { \G} \end{array} \right ] \in\mathbb{Z}_q^{n \times m}$. 
 \item Choose a uniformly random  ${\bf s} \leftarrow \mathbb{Z}_q^{n}$.
\item Sample error vectors $e \leftarrow  D_{\mathbb{Z},\alpha q}, {\e}_{0} \leftarrow  D_{\mathbb{Z},\alpha q}^{\bar{m}}$   and    ${\e}_{1}\leftarrow D_{\mathbb{Z},s'}^{nk}$, where $s'^{2} = (\| {\e}_{0} \| ^ {2} + \bar {m} (\alpha q)^{2}) r^2$. Let the error vector $ {\e} = ({\e}_{0},  {\e}_{1}) \in \mathbb{Z} ^{m}$.
\item Compute $ {\c}_1 = {{\A}}_{id{_i}}^t {\bf s} + {\e} \mod q \in \mathbb{Z}_{q}^{m}$ and  $ {\c}_2 = {\u}_{id_i}^t {\bf s} + e +b\cdot \lfloor q/2 \rfloor  \mod q \in \mathbb{Z}_{q}$.
\item Output the ciphertext $ct = ({\c}_1, {\c}_2) \in\mathbb{Z}_{q}^{m} \times \mathbb{Z}_{q}$.
\end{enumerate}

\paragraph{$\mathbf{Dec}(PP, sk_{id_{i}}, ct)$} On input a public parameter $PP$, the secret key of $i$-th user $sk_{id_{i}}$ and ciphertext $ct$, do:
\begin{enumerate}
\item Compute $b' = {\c}_2 - {\x}_{id_{i}}^t {\c}_1 \in \mathbb{Z}_{q}$.
\item Output $0$ if $b'$ is closer to $0$ than to $\lfloor q/2 \rfloor \mod q$; Otherwise output $1$.
\end{enumerate}

\paragraph{$\mathbf{ReKeyGen}(PP, sk_{id_{i}}, id_i, id_j)$} On input a public parameter $PP$, the secret key of $i$-th user $sk_{id_{i}}$ and identity of $j$-th user $id_j$, do:
\begin{enumerate}
\item Construct ${\A}_{id_i}$ and ${\A}_{id_j}, {\u}_{id_j}$.
\item Choose ${\bf{r}}_{1}\leftarrow  D_{\mathbb{Z}, r}^{{mk}\times n }$ and ${\bf{r}}_{2}\leftarrow D_{\mathbb{Z}, r}^{{mk}\times 1 } $.
\item Construct the proxy re-encryption key  \\ $rk_{i\rightarrow j} = \left [ \begin{array}{c    c  }{\bf{r}}_1 {\A}_{id_j} & ~~~~{\bf{r}}_1{\u}_{id_j} + {\bf{r}}_2 - {P2}({\x}_{id_i} ) \\ {\bf 0}_{1 \times m}  & {\bf{I}}_{1 \times 1} \end{array} \right ]  \in \mathbb{Z} _{q}^{{(mk+1)}\times{(m+1)}}.$
\item Output re-encryption key $rk_{i\rightarrow j}$. 
\end{enumerate}

\paragraph{$\mathbf{ReEnc}( rk_{i\rightarrow j}, ct)$} On input $rk_{i\rightarrow j}$ and $i$-th user's ciphertext $ct = ({\c}_1, {\c}_2)$, do:
\begin{enumerate}
\item Compute the re-encrypted ciphertext $\bar{ct} = (\bar{\c}_1, \bar{\c}_2)$ as follows: \\$\bar{ct}^t = \left [ \begin{array}{c|c} BD({\c}_1)^t & {\c}_2^t \end{array} \right] \cdot rk_{i\rightarrow j} \in \mathbb{Z}_q ^{1 \times (m+1)}.$
\item Output the re-encrypted ciphertext $\bar{ct}$.
\end{enumerate}

\subsection{Correctness and Security}
In this section, we analyze the correctness and security of the proposed scheme. In respect of correctness, the main point is to consider the growth of error due to re-encryption. We have proved that the growth of error is controlled in Theorem \ref{adap_correctness}. Further, we have proved security of the construction in the adaptive identity based model, according to the Definition \ref{def:security}, against chosen plaintext attack in Theorem \ref{thm:adap_security}

\begin{thm} [Correctness] 
\label{adap_correctness}
The $\AIBPRE$ scheme with parameters proposed in Section~\ref{sec:adap_construction}  is correct.
\end{thm}
\noindent{\bf Proof:}
Proof follows from the similar argument of Theorem \ref{sel_correctness}.
\qed

To prove the security, we use the family of abort-resistant hash functions \cite{ABB10-EuroCrypt,BR09,Waters05} $\mathcal{F}_{Wat}$, where  $\mathcal{F}_{Wat}: \{F_h: (\mathbb{Z}_q^l)^* \to \mathbb{Z}_q \}_{h \in \mathbb{Z}_q^l}$ and $F_h(id) = 1 + \sum_{i=1}^{l} h_i b_i$ for $id = (b_1, b_2, \cdots, b_l) \in\{-1, 1\}^l$. $\mathcal{F}_{Wat}$ is a ($Q, \alpha_{min}, \alpha_{max}$) abort-resistant family, where $\alpha_{min} = \frac1{q} (1- \frac{Q}{q})$ by \cite[Lemma 27]{ABB10-EuroCrypt} and $Q$ is number of key extraction query. Since $q \geq 2Q$, we have $\alpha_{min} \geq \frac{1}{2q}$.

\begin{thm}[Security]
\label{thm:adap_security}
The above scheme is $\INDCPA$ secure assuming the hardness of decision-{\em LWE}$_{q, \chi}.$ 
\end{thm}
\noindent{\bf Proof:}
Let the LWE samples of the form $({\bf a}_i, v_i) = ({\bf a}_i, {\bf a}_i^t {\s}+ e_i) \in \mathbb{Z}_q^{n} \times \mathbb{Z}_{q}$, where ${\bf s} \leftarrow \mathbb{Z}_q^{n}$, uniformly random and $e_i \in \mathbb{Z}_{q} $, sample from $\chi$, ${\bf a}_i$ is uniform in $\mathbb{Z}_q^{n}$. we construct column-wise matrix $\A^{*}$ from these samples ${\bf a}_i$ and a vector ${\v}^{*}$ from the corresponding $v_i$.  The proof proceeds in a sequence of games.  Let $\mathbf{W}_i$ be the event that the adversary correctly guessed the challenge bit at the end of $Game~i$. The adversary's advantage in $Game~i$ is $| \Pr[ \mathbf{W}_i] - \frac1{2}| $.

\begin{description}
	\item[Game 0.] This is the original $\INDCPA$ game from definition between an adversary $\mathcal{A}$ against scheme and an $\INDCPA$ challenger.

	\item[Game 1.] In $\bf Game~1$ we change the way that the challenger generates $\bar{\A}, \bar {\A}_i, {\u}_0,$ ${\u}_i$, where $i = 1, \cdots, l $ in the public parameters. In SetUp phase, do as follows:
	\begin{itemize}
	 \item Set the public parameter $ \bar {\A} = \A^{*}$, where $\A^{*}$ is from LWE instance $(\A^{*}, {\v}^*)$.
	 \item Challenger chooses $l$ random scalars $h_i \in \mathbb{Z}_q$ for $i =1, \cdots, l$ and set $\bar {\A}_i=  -{\A}^{*}{\R}_i + h_i {\G} $, where ${\R}_i$ are chosen in the same way as in $Game~ 0$.
	  \item Choose ${\x}_1\leftarrow  D_{\mathbb{Z}, r}^{\bar{m} \times 1}$ and ${\x}_2\leftarrow D_{\mathbb{Z}, r}^{nk \times 1}$; Set ${\x}^* = \left [ \begin{array}{c } {\x}_1 \\  {\x}_2 \end{array} \right ] \in \mathbb{Z}^{m \times 1} $.
	   Set ${\u}_0 = {\A}^{*} \cdot {\x}_1$ and ${\u}_i = - {\A}^{*}{\R}_i \cdot {\x}_2$ for $i =1, \cdots, l$ .
	
\item Set $PP= (\bar{\A}, \bar{\A}_1, \bar{\A}_2, \cdots, \bar{\A}_l, {\u}_0, {\u}_1, \cdots, {\u}_l)$ and send it to the Adversary $\mathcal{A}$.

Due to left-over hash lemma \cite[Lemma 14]{ABB10-EuroCrypt}, the distribution of $({\A}^{*}, -{\A}^{*}{\R}_i,  {\A}^{*} \cdot {\x}_1,  - {\A}^{*}{\R}_i \cdot {\x}_2)$ is statistically indistinguishable with uniform. Hence, $({\A}^{*}, -{\A}^{*}{\R}_i + h_i {\G} ,  {\A}^{*} \cdot {\x}_1,  - {\A}^{*}{\R}_i \cdot {\x}_2)$ is statistically indistinguishable with uniform distribution. So, in adversary's view, these are uniform random matrices, as in ${\bf Game ~0}$. Hence,
\begin{equation}
\label{sec:pre1}
\Pr [W_0] = \Pr[W_1].
\end{equation}
\end{itemize}

\item[Game 2.] In ${\bf Game ~2}$, challenger introduce an abort event which is independent of the Adversary's view. Otherwise, ${\bf Game ~2}$ is identical to ${\bf Game ~1}$. Here, challenger behaves as follows:
\paragraph{{\bf SetUp}} Except  challenger chooses a random hash function $F_h \in \mathcal{F}_{Wat}$, SetUp phase of ${\bf Game ~2}$ is identical with ${\bf Game ~1}$.  Challenger keeps $F_h$ to himself.

\paragraph{{\bf Phase 1}} The adversary $\mathcal{A}$ may make quires polynomially many times in any order to the following oracles:

$\mathcal{O}^{\bf{Extract}}$: To answer a Extract queries on $id_i \in CU$, challenger does the following:
\begin{itemize}
\item If $f_{id_i} =  F_h(id_i) \neq 0$, then using $\bf{Extract}$, Construct\\ ${\A}_{id_{i}}
	= \left [ \begin{array}{c  |  r}{ \A}^* &  -{\A}^* \sum_{j=1} b_j{\R}_j +f_{id_i}\G\end{array} \right ]$ and ${\u}_{id_i} = {\u}_0 + \sum_{j=1}^{l} b_j {\u}_j $.
So, $\sum_{j=1} b_j{\R}_j$ is a trapdoor of ${\A}_{id_{i}}$ with invertible tag $f_{id_i}. \bf{I}$. Then from $\mathbf{Extract}$ algorithm, challenger gets the secret key $sk_{id_{i}}= {\x}_{id_{i}}$ for $id_i$, sends $sk_{id_{i}}$ to the adversary $\mathcal{A}$.\\  For $f_{id_i} = 0$, challenger will abort the game.
\end{itemize}
Challenger will send $\bot$, against the secret key query for $id_i\in HU$.

$\mathcal{O}^{\bf{ReKeyGen}}$: For Re-EncryptionKey queries from $id_i$ to $id_j$, where $id_i, id_j \in CU$ or $id_i \in CU, id_j \in HU$, if $F_h(id_i) \neq 0 \mod q$, then using $\bf{ReKeyGen}$ algorithm, challenger will compute $rk_{i\rightarrow j}$, sends to $\mathcal{A}$. Otherwise, challenger will abort. 
 If challenger answered Extract queries before for $id_i$, then using that keys challenger will construct $rk_{i\rightarrow j}$; otherwise challenger will construct keys as in an extract query first and then use the $\bf{ReKeyGen}$ algorithm to construct rekey, send to $\mathcal{A}$.
 
For Re-EncryptionKey queries from $id_i$ to $id_j$, where $id_i, id_j \in HU$, challenger will construct ${\A}_{id_j}$ and choose ${\bf{r}}_{1}\leftarrow  D_{\mathbb{Z}, r}^{{mk}\times n }$, as in the scheme and one uniformly random matrix $K \leftarrow \mathbb{Z}_{q}^{{mk}\times 1 } $.
It will set the re-encryption key  $rk_{i\rightarrow j} = \left [ \begin{array}{c    c  } {\bf{r}}_1 {\A}_{id_j} & ~~~~K \\ {\bf 0}_{1 \times m}  & {\bf{I}}_{1 \times 1} \end{array} \right ]  \in \mathbb{Z} _{q}^{{(mk+1)}\times{(m+1)}} $, 
send $rk_{i\rightarrow j}$ to the adversary. In adversary's view, simulated {\bf ReKey} and actual {\bf ReKey} are identical.

$\mathcal{O}^{\bf{ReEnc}}$: For re-encryption query challenger first check that there was Rekey query before or not, if yes, then challenger will do as follows: 
\begin{itemize}
\item If there is a ReEncKey, using that ReEncKey, challenger will compute $\mathbf{ReEnc}(rk_{i \rightarrow j}, ct)$ according to the $\mathbf{ReEnc}$ algorithm to reply the adversary. 
\item If there was an abort, Challenger will abort here, too.
\end{itemize}
 Otherwise, challenger will compute ReEnc Key (following the criteria for $\mathcal{O}^{\bf{ReKeyGen}}$) maintain the restrictions as in definition \ref{def:security} and computes $\mathbf{ReEnc}(rk_{i \rightarrow j}, ct)$ according to the $\mathbf{ReEnc}$ algorithm to reply the adversary.

\paragraph{{\bf Challenge}} In the challenge phase, the challenger checks if the challenge identity ${id{_{i^*}}} (\in HU) = (b_1^*, \cdots, b_l^*)$ satisfies $f_{id{_{i^*}}} = 1 + \sum_{j=1}^{l} h_jb_j^* = 0 $. If not, challenger abort the game ( and pretends that the adversary output a random bit $r'$ in $\{0, 1\}$ in Decision Phase), challenger will produce the challenge ciphertext ${ct}^* = ({\c}_1^*, {\c}_2^*)$ for the challenge identity $id_{i^*}$ on a message  ${{b}^*}\in \{0,1\}$. Since, $f_{id{_{i^*}}} = 0$, so $ {\A}_{id{_{i^*}}}= \left [ \begin{array}{c  |  r} {\A}^*  &  -{\A}^* \sum_{j=1}^{l} b_j^* {\R}_j \end{array} \right ]$. Let ${\R}^* = {\sum_{j=1}^{l} b_j^* {\R}_j }$. We will treat  ${\x}^* = \left [ \begin{array}{c } {\x}_1 \\  {\x}_2 \end{array} \right ]$ as secret key of ${id{_{i^*}}}$. We will use ${\x}^*$ to answer ReKey queries from ${id{_{i^*}}}$  in $Phase~2$. Check that ${\A}_{id{_{i^*}}} \cdot {\x}^* = {\u}_0 + \sum_{j=1}^{l} b_j ^*{\u}_j$ i.e. ${\A}_{id{_{i^*}}} \cdot {\x}^*= {\u}_{id{_{i^*}}}$.

Challenger will produce the challenge ciphertext ${ct}^* = ({\c}_1^*, {\c}_2^*)$ for the challenge identity $id_{i^*}$ on a message  ${{b}^*}\in \{0,1\}$ as follows: 
\begin{itemize} 
\item Choose a uniformly random  ${\bf s} \leftarrow \mathbb{Z}_q^{n}$ and noise vectors $  {\e}_{0} \leftarrow  D_{\mathbb{Z},\alpha q}^{\bar{m}}$.

  Set $ {\c}_1^* =  \left [ \begin{array}{c}{\v}^{*} \\ -{{\R}^*}^t {\v}^{*} \end{array} \right ]  \in \mathbb{Z}_{q}^{m}$  and $ {\c}_2^* =  {{\x}^*}^t  \cdot \left [ \begin{array}{c} {\v}^{*} \\ -{{\R}^*}^t {\v}^{*} \end{array} \right ] +b^{*} \cdot \lfloor q/2 \rfloor  \in \mathbb{Z}_{q}$, where ${\v}^{*} = {{\A}^{*}}^t \cdot{\s} + {\e}_0$. Send $ct^* = ({\c}_1^*, {\c}_2^*)$ to the adversary.
 \item Here ${\c}_1^*, {\c}_2^*$ satisfies,  
  \begin{align*}
	{\c}_1^*  &=  \left [ \begin{array}{c}{\v}^{*} \\ -{{\R}^*}^t {\v}^{*} \end{array} \right ]\\
	              &=   \left [ \begin{array}{c} {\A}^{*t} \cdot{\s} + {\e}_0 \\ -{{\R}^*}^t {\A}^{*t} \cdot{\s} - {{\R}^*}^t {\e}_0 \end{array}\right ] \\
	               &=  \left [ \begin{array}{c} {\A}^{*t}  \\ -{{\R}^*}^t {\A}^{*t} \end{array} \right ] \cdot{\s} +  \left [ \begin{array}{c} {\e}_0 \\ - {{\R}^*}^t {\e}_0  \end{array} \right ] \\
	              &=  {{\A}}_{id{_{i^*}}}^t {\bf s} +  \left [ \begin{array}{c} {\e}_0 \\ - {{\R}^*}^t {\e}_0 \end{array} \right ]  \in \mathbb{Z}_{q}^{m}
	\end{align*}

 \begin{align*}
	{\c}_2^*  &= {{\x}^*}^t  \cdot \left [ \begin{array}{c} {\v}^{*} \\ -{{\R}^*}^t {\v}^{*} \end{array} \right ] +b^{*} \cdot \lfloor q/2 \rfloor \\
	&= {{\x}^*}^t  \cdot \left [ \begin{array}{c} {\A}^{*t} \cdot{\s} + {\e}_0 \\ -{{\R}^*}^t {\A}^{*t} \cdot{\s} - {{\R}^*}^t {\e}_0 \end{array}\right ] +b^{*} \cdot \lfloor q/2 \rfloor\\
        &= {{\x}^*}^t  \cdot {{\A}}_{id{_{i^*}}}^t {\bf s} + {{\x}^*}^t  \left [ \begin{array}{c} {\e}_0 \\ - {{\R}^*}^t {\e}_0  \end{array} \right ] +b^{*} \cdot \lfloor q/2 \rfloor\\
	&= {\u}_{id{_{i^*}}}^t {\s} + {{\x}^*}^t  \left [ \begin{array}{c} {\e}_0 \\ - {{\R}^*}^t {\e}_0  \end{array} \right ] +b^{*} \cdot \lfloor q/2 \rfloor \in \mathbb{Z}_{q}
\end{align*}
\end{itemize}
By Corollary 3.10 in \cite{Regev05}, the noise term $( {\e}_0, - {{\R}^*}^t {\e}_0)$ of ${\c}^*_1 $ is within $ \negl$ statistical distance from discrete Gaussian distribution $D_{\mathbb{Z},s'}^{nk}$. The same argument, also, applies for the noise term of ${\c}^*_2$. So, $({\c}_1^*, {\c}_2^*)$ is the valid challenge ciphertext in ${\bf Game~2}$\cite[Lemma 12, Lemma 14]{ABB10-EuroCrypt}.

\paragraph{{\bf Phase 2}} After receiving the challenge ciphertext, $\mathcal{A}$ continues to have access to the $\mathcal{O}^{\bf{Extract}}$, $\mathcal{O}^{\bf{ReKeyGen}}$ and $\mathcal{O}^{\bf{ReEnc}}$ oracle as in {\bf Phase 1}. But for ReEnc Key query from the challenge identity ${id{_{i^*}}}$ to any honest user $HU$, challenger use ${\x}^* = \left [ \begin{array}{c } {\x}_1 \\  {\x}_2 \end{array} \right ]$ as the secret key of $id^*$, then using the ${\bf{ReKeyGen}}$ algorithm, it constructs the Rekey and send to the adversary.

\paragraph{{\bf Decision}} Note that the adversary never sees $F_h$ and has no idea if an abort event took place. It is convenient to consider this abort at the Decision Phase. Nothing would change if the challenger aborted the game as soon as the abort condition became true. Let $id_1, id_2, \cdots, id_Q \in CU$ be the identities, on which adversary did Extract and ReKey queries.

In the final guess phase, the adversary outputs its guess $r' \in \{0, 1\}$ for $r$. The challenger now does the following:
\begin{enumerate}
\item $\bf{Abort~check}$: The challenger checks if $F_h({id{_{i^*}}}) = 0 $ and $F_h(id_i) \neq 0$ for $i = 1, \cdots, Q$. If not, it overwrites $r'$ with a fresh random bit in $\{0, 1\}$ and we say that challenger aborted the game. Note that the adversary never sees $F_h$ and has no idea if an abort event took place. 
\item $\bf{Artificial~Abort}$: The challenger samples a bit $\Gamma \in \{0, 1\}$ such that $\Pr[ \Gamma = 1] = \gamma ({id{_{i^*}}}, id_1,$ $\cdots, id_Q)$, where the function $\gamma (\cdot)$ is defind in \cite[Lemma 28]{ABB10-EuroCrypt}. If $\Gamma = 1$ the challenger overwrites $r'$ with a fresh random bit in $\{0, 1\}$ and we say that challenger aborted the game due to an artificial abort; see \cite{ABB10-EuroCrypt} for more details.
\end{enumerate}

This completes the description of ${\bf Game ~2}$. Note that the abort condition is determined using a hash function $F_h$ that is independent of the adversary’s view. A similar argument as in \cite[Theorem 25]{ABB10-EuroCrypt} yields that 
\begin{equation}
\label{sec:pre2}
| \Pr[ \mathbf{W}_2] - \frac1{2}| \geq \frac{1}{4q} | \Pr[ \mathbf{W}_1] - \frac1{2}|.
\end{equation}



\item[Game 3.] Here, we only change how the ${\v}^{*}$ component of the challenge ciphertext is created, letting it be uniformly random in $\mathbb{Z}_{2q} ^{\bar m}$. Challenger construct the public parameters, answer the secret key queries, re-encryption queries and construct the remaining part of the challenge ciphertext exactly as in ${\bf Game ~2}$. It follows from the hardness of the decisional LWE$_{q, \chi}$ that ${\bf Game ~2}$ and ${\bf Game~3}$ are computationally indistinguishable.

Now, by the left-over hash lemma \cite[Lemma 14]{ABB10-EuroCrypt}, (${ \A}^*,{\v}^{*}, -{ \A}^{*}{\R}^*,-{{\R}^*}^t{{\v}^{*}}$) is $\negl$-uniform where ${ \R}^*$ is same as in ${\bf Game ~2}$. Therefore, the challenge ciphertext has the same distribution (up to  $\negl$ statistical distance) for any encrypted message. So, the advantage of the adversary against the proposed scheme is same as the advantage of the attacker against decisional LWE$_{q, \chi}$. Since $\Pr[W_3] = \frac 1{2}$, we obtain 
\begin{equation}
\label{sec:pre4}
|\Pr[ W_2 ] - \frac 1{2}| = |\Pr[W_2] - \Pr[W_3]| \leq \epsilon.
\end{equation}
\end{description} 
From equation (\ref{sec:pre1}), (\ref{sec:pre2}) and (\ref{sec:pre4}), we get $|\Pr[ W_0 ] - \frac 1{2}| \leq 4q \epsilon.$ This completes the proof.
\qed

\section{Conclusion}
In this paper, we first propose quantum-safe concrete constructions of collusion-resistant\\ $\SIBPRE$ and $\AIBPRE$ secure in standard model. All the proposed constructions are single-hop. It is an interesting open issue to construct multi-hop version of the proposed schemes.

\bibliography{latbib}
\bibliographystyle{splncs04}

\end{document}